\documentclass[]{article}

\usepackage[a4paper,left=2cm,right=2cm,top=2cm,bottom=4cm,bindingoffset=5mm]{geometry}
\usepackage[pdftex]{graphicx}
\title{Comet 12P/Pons-Brooks:  Identification with Comets C/1385 U1 and C/1457 A1}
\author{Maik Meyer$^{1}$, Takao Kobayashi$^{2}$, Syuichi Nakano$^{3}$,	Daniel W. E. Green$^{4}$
\\
\\
$^{1}$ Limburg, Germany\\
$^{2}$ Oizumi, Gunma-ken, Japan\\
$^{3}$ Sumoto, Japan\\
$^{4}$ Harvard University, Cambridge, MA, USA
}
\begin{document}

\maketitle
\begin{center}
	\textit{Submitted and to be published in the International Comet Quarterly}
\end{center}
\begin{abstract}
Comet 12P/Pons-Brooks is an intermediate-period comet (orbital period around 71 years) comparable to 1P/Halley.  The comet was seen first in 1812, and then again in 1883 and 1954. The comet was recovered in 2020 an will pass its perihelion passage 2024.  We report the unambiguous identification of this comet with the historic comets C/1385 U1 and C/1457 A1; we were able to link both historic apparitions with 12P.  The link is also supported by the historic descriptions of its appearance and brightness.  Further, we discuss other historical comets for possible relationships with 12P and identify a comet seen in 245 as a probable earliest recorded sighting of 12P/Pons-Brooks.
\end{abstract}

\section{Introduction}

Periodic comet 12P/Pons-Brooks was discovered by J. L. Pons on 1812 July 21 from Marseille, France; the comet was followed during that apparition until September 28. It was re-discovered by W. R. Brooks on 1883 September 2 from Phelps, NY, USA, and followed until 1884 June 2; during that apparition, the comet experienced several outbursts.  At its next apparition, in 1954, comet 12P also exhibited several outbursts.  Condensed observational details of all three apparitions can be found in Kronk (2003, 2009).

Despite the fairly high absolute magnitude of around 4-5 and the comet's apparent tendency to have occasional outbursts, it seems that no searches for earlier appearances have been made using historical data.  In February 2020, the first author integrated the orbit of 12P backward until about the year 1000.  The calculations used data from the apparitions of 1883-1884 and 1953-1954 (taken from the Minor Planet Center's online database).  From the integration backwards, it was apparent that the orbit for this comet is very stable and does not experience strong planetary gravitational perturbations in the covered period.  From a check of different cometographies, it was apparent that the first comet of 1457 and the comet of 1385 were almost perfect matches concerning the perihelion time.  As a next step, these backward-integrated orbits were compared with catalogued orbits for the 1385 and 1457 comets, using a planetarium software program (``GUIDE", by B. Gray; cf. website URL {\tt https://www.projectpluto.com}) that showed that the integrated orbits for 12P were fully compatible with the observed paths and the observational circumstances of the 1457 and 1385 comets. Not only did 12P appear positioned within the area indicated by the ancient observations but also the sense of movement did fit perfectly. By adjusting the perihelion time by a few days for each of these apparitions, the match could be brought even closer.

\section{The Comet of 1457}

In 1864 a manuscript by Italian cartographer and astronomer P. Toscanelli was found in the National Library in Florence, Italy.  The manuscripts contained observations of six comets seen by Toscanelli in the 15th century, with their positions drawn into celestial maps drawn by Toscanelli.  After the discovery of the manuscripts, G. Celoria performed in-depth investigations of Toscanelli's manuscripts and derived orbits from the positions drawn by Toscanelli.  The first comet of 1457 was observed daily on five nights between 1457 January 23 and 27.  Celoria published his analyses, including the derived orbits, several times (Celoria 1884, 1894, 1921).  His orbit is based on three of the five observations and via comparison with the other two observations. He correctly states that the orbit is not of much accuracy due to the very short arc.

The Toscanelli drawings show a short tail extending to about $0.5{\circ}$. There are indications of a coordinate grid in Toscanelli's drawings that at times seems incomplete, but it helped Celoria to identify the area of the sky where the comet was seen. A more contemprorary analysis of Toscanelli's maps can be found in Jervis (1985). Figures 1 and 2 show the original drawing by Toscanelli and the representation by Celoria, respectively.  It can be seen from the images that the accuracy can only be good to maybe a few degrees, and any orbit derived from them is prone to some considerable uncertainty, something that Celoria (1894) acknowledged quite clearly:  ``The observations, despite their nature, are quite well represented from my orbital elements, but they, even if of a remarkable precision for that time, are too close to each other and too few to judge with certainty how much the orbital elements themselves are close to the real ones."\\

\begin{center}
	\includegraphics[width=0.8\textwidth]{"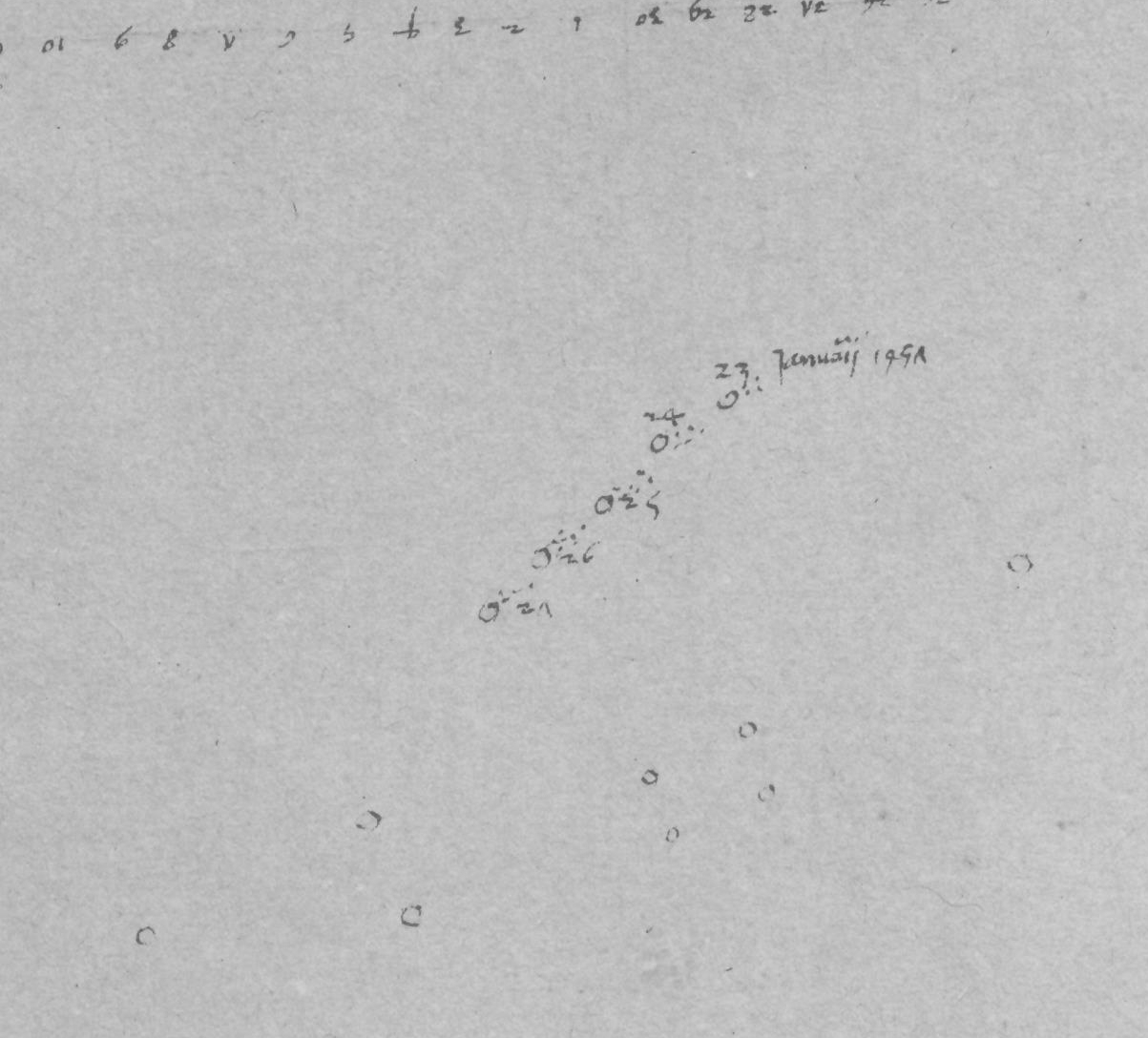"}\\
\end{center}

Figure 1: Drawing of the positions of the comet of January 1457 by P. Toscanelli. Image courtesy F. Stoppa, Milan, Italy.\\

But Toscanelli was apparently not the only observer of this comet.  There exists a Chinese observation of the 1457 comet that, however, bears a problem.  The Chinese reports are as follows (Pankenier, Xu, and Jiang 2008) and are given for 1457 January 14:  ``7th year of the Jingtai reign period of Emperor Yingzong of the Ming Dynasty, 12th month, day jiayin [51], at night, a broom star with bright rays 5 cun long reappeared in lunar mansion BI [LM 19], slowly traveling southeastward.  Its bright rays gradually lengthened from this day through day guihai [60].  [\textit{Ming Yingzong shilu}] ch. 273".  Another Chinese source given there (\textit{Ming shi: tianwen zhi}, ch. 27) supplies basically the same information.\\

\begin{center}
	\includegraphics[width=0.80\textwidth]{"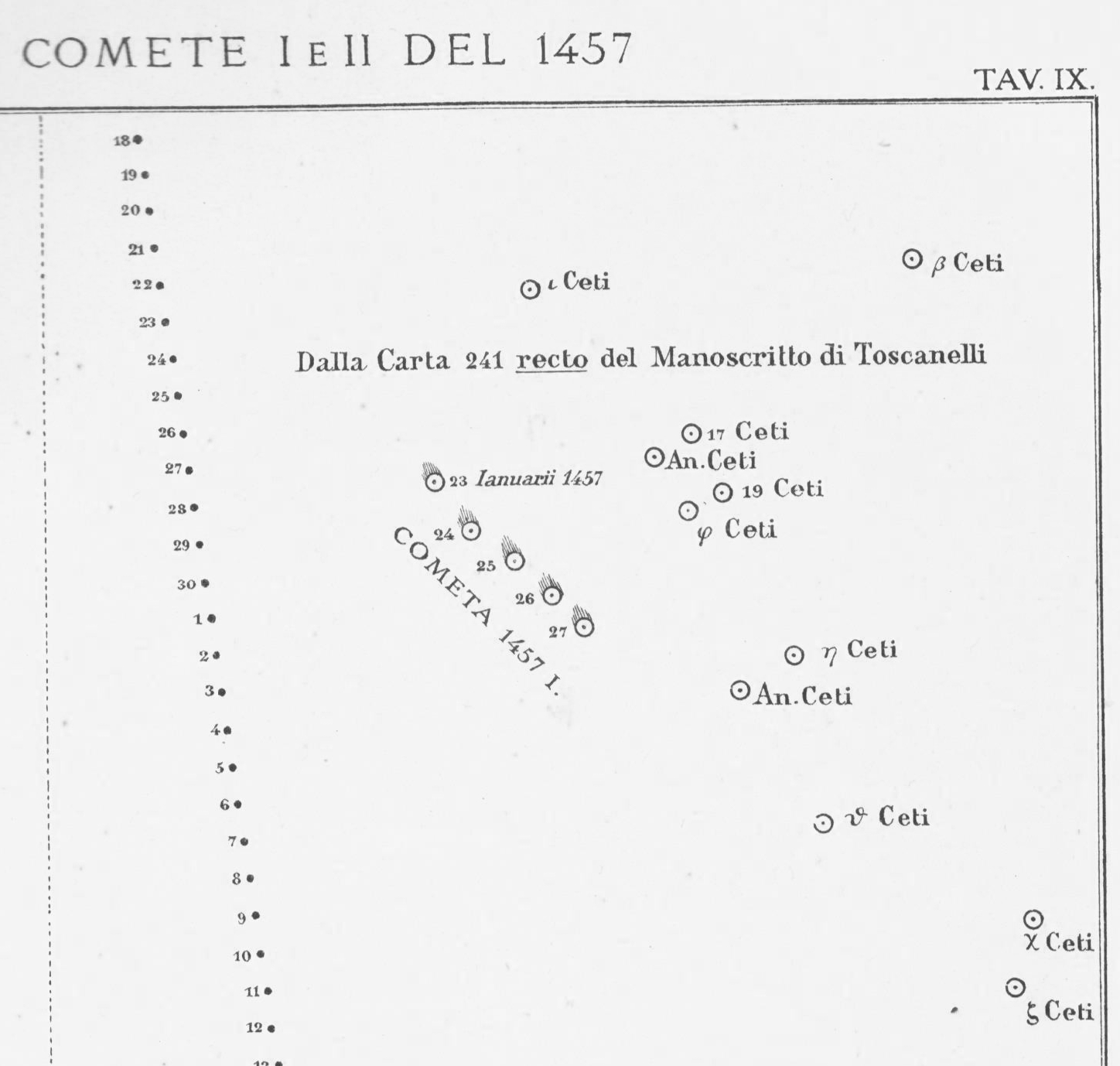"}\\
\end{center}

Figure 2: Representation of Toscanelli's drawing by G. Celoria (1894). Image courtesy F. Stoppa, Milan, Italy.\\

Stryuck (1740, p.\ 247) wrote:  ``Besides the story about the Comet one finds in the last mentioned Chronicle [{\it Anton.\ de Ripalta Annal.\ Placent., col.\ 905}] that in the Year 1456, in the Month of December, and in the Year 1457, in January, four strange Stars appeared, moving from the East to the West, almost in the shape of a Cross.  This could easily be some Fixed Stars or one or more Planets.  {\it Ludov.\ Cavitel.\ Cremonen.\ Annal.}\ (col.\ 1456, tom.\ 3, par.\ 2) tells about a Comet that was seen in the Year 1456, the 5th of December, and another one in January ..." Interestingly, both Struyck and Lubienietz (1667) mention the comets of June 1456 and June 1457, but nothing more on the comet of Jan. 1457 (other than that above).  And neither of these two cometographers mentioned the 1385 comet at all; as they both had access to many European historical materials on comets, it says a lot that these comets were not apparently widely known in Europe, so they must not have been very bright and thus not easily seen from Europe.

What can be deduced from the Chinese texts?  First, the object was seen as early as 1457 January 14 -- but possibly earlier, since the word `re-appeared' is used. Second, the tail was about $0.5^{\circ}$ long; 1 cun is about $0.1^{\circ}$.  This nicely agrees with the length derived by Celoria from Toscanelli's drawings. Third, the object was seen for about 10 days, until January 23.

The main problem with the Chinese observation (which was also discussed at length by Celoria) is the position in the lunar mansion Bi (or Pi), which corresponds to lunar mansion 19 and refers to the area around the Hyades and $\varepsilon$ Tau, also called the ``Hunting Net".  However, using the orbital elements of 12P, the comet would have been near $\omega$ Psc and $\gamma$ Peg on that date.  It can be concluded that the Chinese position is not at all compatible with the Toscanelli observations, except there is an error in the Chinese sources (see also the similar remark by Celoria).  For the Chinese comet, 12P would have been in the 14th lunar mansion, called Dongbi or Tung-bi but sometimes also given as Pi (Ho 1962).  If the 14th and 19th lunar mansions were mixed up, then 12P would be correctly placed on 1457 January 14; but, of course, this cannot be proven anymore.
	
Finally, another argument for the identity with 12P:  Based on the derived magnitude parameters of the apparition of 1953, the magnitude of 12P in 1457 was perhaps at mag 3-4 (assuming no outburst), with the comet being close to perihelion and about 0.95 AU from the earth.  This would explain why it was not such a conspicuous object.  If Celoria's orbit were correct, the comet would have brightened further with increasing elongation in the following days which raises the question why it was not observed further by Toscanelli. The orbit of comet 12P shows that it in fact became fainter with a slowly increasing elongation, which explains the short observation period.
	
As a sidenote, it should be mentioned that the 1457 comet was long suspected to be identical with what later became known as comet 27P/Crommelin (Schulhof 1885; Galle 1894, p.\ 157; Procter and Crommelin 1937).  Modern calculations were not able to confirm this and, moreover, that identity can be ruled out (Marsden 1975; Festou, Morando, and Rocher 1985).
	\\
	
The orbit by Celoria is given here for reference
	\\
	
$ 	q = 0.703\ AU, peri = 195^\circ , node = 258^\circ , i = 13^\circ   $
	\\
	
\section{The Comet of 1385}  
	
For the 1385 apparition, we only have the description of the comet's apparent movement from Asian sources:  the \textit{Ming Taizu shilu}, ch. 175, and \textit{Ming shi: tianwen zhi}, ch. 27 (Pankenier et. al. 2008).  On 1385 October 23, the comet appeared near Coma Berenices, Leo, and Virgo, and	after that, moved towards $\beta$ Vir and left the area of $\beta$ and $\eta$ Vir.  On 1385 October 30, the comet entered Crater; on November 4 it 'trespassed against' an asterism in Hydra.  The comet had a 10-degree tail according to Biot (1843a; see also Carl 1864, p.\ 42).  The widely cited orbit by Hasegawa (1979) of course resembles this general movement.
	
The orbit of 12P is perfectly consistent with the above description and moves similar to comet C/1385 U1; the apparent path in the sky fits the description from the Chinese records even better.  Using the magnitude parameters from the	1953 apparition the brightness was perhaps around magnitude 2 (assuming no outburst), since the	apparition was very favorable due to a close approach to the earth.  This agrees well with the Chinese observations, too.
	\\
	
Several orbits have been calculated from the Chinese descriptions in the past. The orbits (2000.0) derived by Peirce (1846), Hind (1846) and Hasegawa  are given here for reference.\\

	Peirce: 	$q = 0.755\ AU, \omega = 155^\circ, \Omega = 270^\circ, i = 105^\circ
	$
		
	Hind:		$q = 0.738\ AU, \omega = 130^\circ, \Omega = 296^\circ, i =  52^\circ
	$
		
	Hasegawa:	$q = 0.79\  AU, \omega = 289^\circ, \Omega = 103^\circ, i = 103^\circ
	$
	\\
	
As noted above for the 1457 comet, neither Struyck nor Lubienietz mentioned the 1385 comet at all; since they both had access to many European historical materials on comets, it says a lot that these comets were not apparently widely known in Europe, so they must not have been very bright and thus not easily seen from Europe.
	
Nevertheless, the comet was not completely missed in Europe. J. Meyerus	Baliolanus (1561) gives an account of a comet seen on the feast day of Saints Cosmas and Damian, which corresponds to September 27, 1385, when the comet would have been at magnitude 5, but it may have been in outburst then. However,	his text also says that the comet appeared in October. He goes on to say that the comet did shine in many colours. A similar account can be found in a later annal by E. Sueyro (1624) with the only difference that he put it in the year 1386. From the description it could also relate to an aurora. Another mention	can be found in the annals the German city of Trier (1838). The editors of this edition remark on a manuscript which states that in 1385 a terrible comet 	appeared. The comet can also be found in annals of Iceland (1847) which simply says for 1385 that a comet appeared.
	
\section{Linkage}  
	
For the linkage of the apparitions of 1385 and 1457, the following positions were derived from the descriptions in the historical sources.\\
	
\begin{verbatim}
	1385 UT             R.A. (2000) Decl.       Mag.
       Oct. 22.9       12 00         +12 00         2
            29.9       11 55         - 8 00
       Nov.  3.9       11 55         -35 00

   1457 UT             R.A. (2000) Decl.       Mag.      Observer
      Jan. 23.7        0 40         - 3 40         3     Toscanelli
           25.7        0 47         - 5 20                 "
           27.7        0 56         - 7 15                 "
\end{verbatim}	
	
The observations of the apparitions of 1812 and 1883-1884 were re-reduced recently by co-author T. Kobayashi, from which a linked orbit could be derived and which was published in Green (2020a) and Nakano (2020a).
	
Prompted by these announcements, the comet was recovered on June 10 and 17, 2020, with the Lowell Observatory 4.3-m Discovery Telescope and the Large Monolithic Imager by Ye et. al. (2020a) with the comet at a distance of 11.9 AU. On stacked images a broad tail of 3' length was visible implying that it was already active. The correction to the orbit by Kobayashi based on the data from 1385 - 1954 was only +0.16 day (see Green (2020b).
	
Including these recovery observations the following linked orbital elements were derived.  These observations are listed in Table 1. His elements are based on a total of 1052 astrometric observations and include perturbations by Mercury-Neptune and Ceres, Pallas, and Vesta. Non-gravitational effects were included in the orbit computation. The weighted mean residual is 1$''$\llap.41.  The comet passed 3.71 AU from Uranus on 1819 Apr. 26 and 1.62 AU from Saturn on 1957 July 29 UT.  The comet has made numerous	close approaches to the earth (0.41 AU on 1385 Oct. 29, 0.90 AU on 1457 Jan. 10, and 0.63 AU on 1884 Jan. 9 UT).
	
It should be noted that a correction in ET-UT for the 1385 and 1457 observations was ignored since no definitive values for ET-UT are available.  However, following Stephenson (1997) and using approximate values for (ET - UT) of +330	s for 1385 and +220 s for 1457, the residuals amount to about 48$''$ and 12$''$, and perihelion time corrections of only about -0.002 and +0.005 day, respectively.
\\
\begin{verbatim}

	Epoch = 1385 Nov.  8.0 TT
	T = 1385 Nov.  6.327 TT          Peri. = 200.036
	e = 0.95505                      Node  = 255.125   2000.0
	q = 0.78362 AU                   Incl. =  73.829
	a = 17.431967 AU     n = 0.013542     P =  72.78 years
	
	Epoch = 1457 Jan. 14.0 TT
	T = 1457 Jan. 30.1002 TT         Peri. = 199.9041
	e = 0.954800                     Node  = 255.2502  2000.0
	q = 0.778438 AU                  Incl. =  74.0399
	a = 17.22216 AU    n = 0.0137903    P =  71.47 years
	
	Epoch = 1812 Aug. 30.0 TT
	T = 1812 Sep. 15.82612 TT        Peri. = 199.29022
	e = 0.9553274                    Node  = 255.63879 2000.0
	q = 0.7771051 AU                 Incl. =  73.95643
	a = 17.3955643 AU   n = 0.01358458   P =  72.55 years
	
	Epoch = 1884 Jan. 25.0 TT
	T = 1884 Jan. 26.21681 TT        Peri. = 199.17679
	e = 0.9550368                    Node  = 255.77454 2000.0
	q = 0.7757320 AU                 Incl. =  74.04048
	a = 17.2526163 AU   n = 0.01375377   P =  71.66 years
	
	Epoch = 1954 May  18.0 TT
	T = 1954 May  22.88058 TT        Peri. = 199.02746
	e = 0.9548317                    Node  = 255.89097 2000.0
	q = 0.7736564 AU                 Incl. =  74.17689
	a = 17.1283021 AU   n = 0.01390377   P =  70.89 years
	
	Epoch = 2024 May  10.0 TT
	T = 2024 Apr. 20.99698 TT        Peri. = 198.98718
	e = 0.9545914                    Node  = 255.85595 2000.0
	q = 0.7807641 AU                 Incl. =  74.19138
	a = 17.1941867 AU   n = 0.01382393   P =  71.30 years
\end{verbatim}	
	
Figures 3 and 4 show the apparent paths for both apparitions based on Celoria's	and Hasegawa's orbits, respectively.  They also show the paths based on the linked orbit by Kobayashi.  It can be seen that the apparent paths from the linked orbit by Kobayashi are quite similar to the apparent paths from orbits	by both Celoria and Hasegwawa.\\
		
\begin{center}
	\includegraphics[width=0.9\textwidth]{"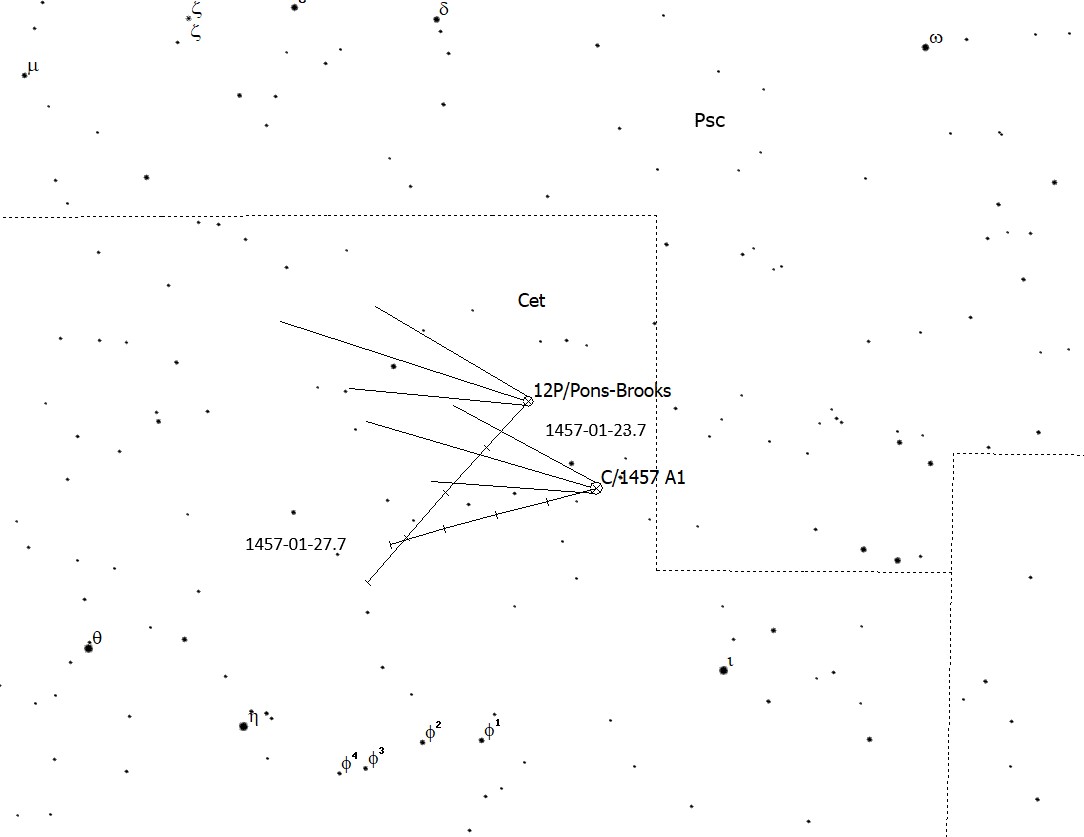"}\\
\end{center}

Figure 3: Comparison of the paths of comet C/1457 A1 based on Celoria's orbit and comet 12P/Pons-Brooks based on the linked orbit.\\

\begin{center}
	\includegraphics[width=0.9\textwidth]{"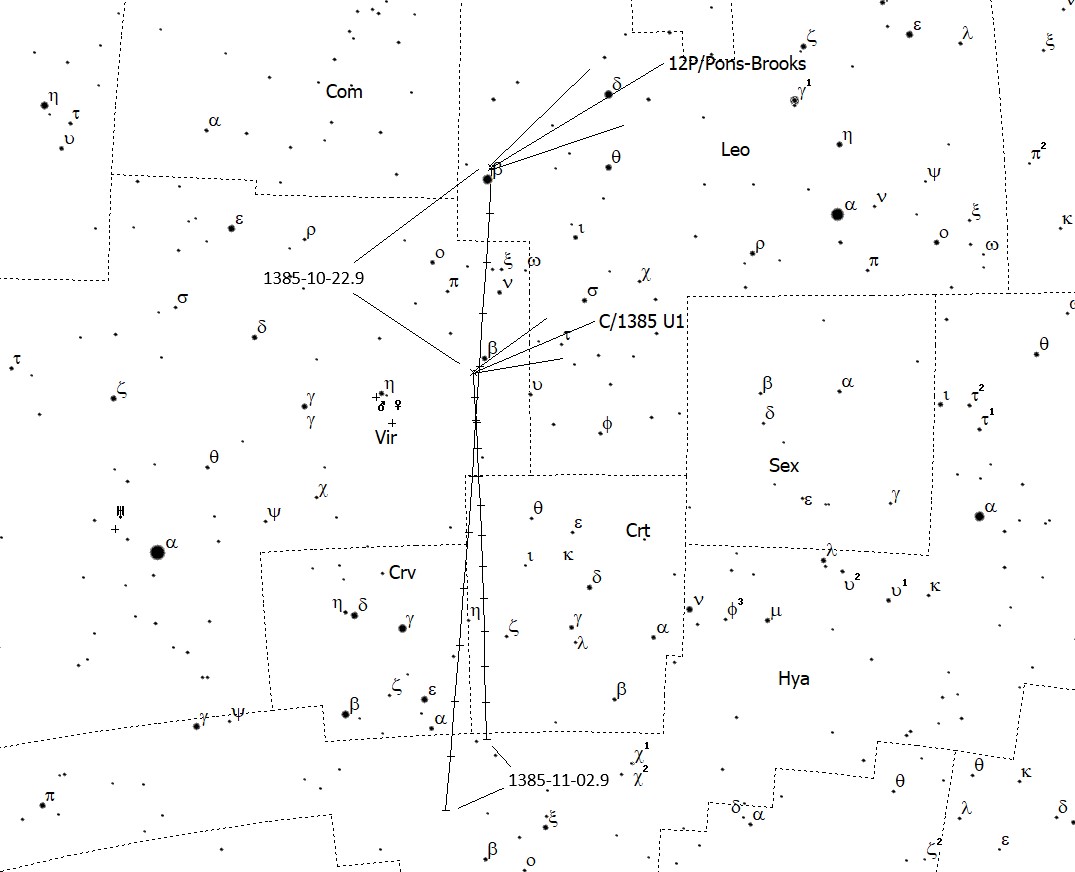"}\\
\end{center}

Figure 4: Comparison of the paths of comet C/1385 U1 based on Hasegawa's orbit and comet 12P/Pons-Brooks based on the linked orbit.\\
	
\section{Discussion of sightings at other apparitions}
	
On the basis of the orbits given in Table 2 below, a search was conducted in historical comet reports for other sightings of comet 12P/Pons-Brooks. It is reasonable to assume that non-gravitational forces, which are also present for this comet, should not change the predicted orbits before	1385 by a large amount, since this would require a substantial change in these forces that were quite constant between 1385 and now.
	
It should be stressed that for a comet to be noticed by chance without a telescope, it needs to be placed in dark skies at solar elongations > $40^{\circ}-50^{\circ}$, and at a certain brightness (say, brighter than visual magnitude 3-4 in a moonless sky).  For most of the apparitions discussed in the text below,	this is not the case.  The apparitions of 1457 and 1385 were very	favorable, where the comet was close to the earth and bright enough to be easily seen.  From the orbital circumstances, a perihelion occurring between August and January provides the most favorable viewing conditions. However, one has to take into account that 12P is prone to outbursts. This is why it seems nevertheless useful to look at each apparition and see whether other historic candidates are available.\\
	
	Table 2:  Orbital elements by T. Kobayashi for 12P/Pons-Brooks, based on observations from 1385-2020.
	
\begin{small}
\begin{verbatim}
	0012     33 03 24.392  0.783058  0.955528 201.801  253.835   73.443    330331
	0012    105 08 27.621  0.777178  0.955319 201.719  253.956   73.656   1050820
	0012    176 05 17.343  0.774058  0.954814 201.597  254.078   73.899   1760519
	0012    245 09 15.466  0.772882  0.954562 201.409  254.157   74.028   2450914
	0012    315 03 12.893  0.785689  0.954230 201.456  254.037   73.992   3150331
	0012    386 09 08.388  0.787368  0.954884 201.372  254.025   73.741   3860825
	0012    459 07 16.245  0.787975  0.955172 201.303  254.095   73.647   4590802
	0012    532 06 28.502  0.783021  0.955135 201.252  254.266   73.771   5320709
	0012    603 07 03.510  0.777660  0.954684 201.134  254.378   74.057   6030627
	0012    673 01 08.923  0.777703  0.954382 200.940  254.435   74.145   6730110
	0012    742 10 28.225  0.779050  0.954594 200.797  254.473   74.053   7421015
	0012    813 08 06.236  0.780271  0.955070 200.671  254.407   73.909   8130823
	0012    886 04 02.499  0.782316  0.955414 200.607  254.504   73.748   8860402
	0012    959 01 24.841  0.777944  0.955474 200.548  254.627   73.733   9590129
	0012   1030 07 12.737  0.772739  0.955157 200.430  254.763   73.951  10300625
	0012   1100 06 12.277  0.770363  0.954827 200.273  254.872   74.143  11000617
	0012   1170 02 27.511  0.781288  0.954490 200.288  254.843   74.039  11700210
	0012   1241 04 11.464  0.783938  0.954852 200.192  254.899   73.886  12410418
	0012   1313 05 01.651  0.784435  0.955108 200.093  254.962   73.812  13130510
	0012   1385 11 06.327  0.783616  0.955047 200.036  255.125   73.829  13851108
	0012   1457 01 30.100  0.778438  0.954800 199.904  255.250   74.040  14570114
	0012   1527 03 12.625  0.776586  0.954673 199.720  255.326   74.106  15270328
	0012   1597 07 03.752  0.777688  0.954791 199.582  255.410   74.060  15970618
	0012   1668 04 17.408  0.777538  0.955174 199.441  255.389   73.979  16680426
	0012   1740 07 14.017  0.779445  0.955366 199.371  255.513   73.900  17400628
	0012   1812 09 15.826  0.777105  0.955327 199.290  255.639   73.956  18120830
	0012   1884 01 26.217  0.775732  0.955037 199.177  255.775   74.040  18840125
	0012   1954 05 22.881  0.773656  0.954832 199.027  255.891   74.177  19540518
	0012   2024 04 20.997  0.780764  0.954591 198.987  255.856   74.191  20240510
	
\end{verbatim}	
\end{small}	
In the following we discuss earlier returns to perihelion that were missed by observers.  Returns to perihelion prior 1385 are only discussed if there was a promising candidate identified or the comet experienced a particularly favorable apparition.
	\\
	
	\textbf{Perihelion 1740 July 14}
	\\
This apparition was not favorable concerning the observing geometry.  The comet	might have become brighter than magnitude 10 in April, but already at an elongation below $50^{\circ}$. In May the brightness may have attained magnitude 7, but the elongation was then below $40^{\circ}$.  Perihelion was reached almost behind the sun, the elongation being then around $16{\circ}$, the brightness perhaps magnitude 4. The comet then moved quickly southward and remained at elongations below $45^{\circ}$. There is also no promising candidate in any known records to be found.
	\\
	
	\textbf{Perihelion 1668 April 17}
	\\
This return to perihelion was also not favorable, with the observing geometry being very poor.  In January the comet might have been at magnitude 9 at elongations of just below $60^{\circ}$. Perihelion was reached at only $22^{\circ}$ elongation with a magnitude of perhaps 4.
	
There was one comet observed in 1668 from March 3 to 30, which is known as C/1668 E1.  Its orbit is in no way compatible with 12P and can clearly be excluded.  This was a very bright comet, a sunskirter with a perihelion distance of only 0.066 AU that had a long tail and was brighter than Venus.
	
There are other records of a comet seen earlier in 1668. In a paper by Park and Chae (2007), a comet is mentioned that was seen by Korean observers from March 11. While Park and Chae attribute this object to 12P, it is more likely that it is another description of C/1668 E1.  It would have been strange to see another bright comet in the same general region of the sky, as only one bright known comet was then observed widely throughout the world.	And it again has to be stressed that 12P was likely near magnitude 11
on 1668 March 11!  It would have taken a very large and long-lived outburst	to bring it to a brightness level to be seen with the unaided eye by the Koreans (and then only by the Koreans).
	\\
		
	\textbf{Perihelion 1597 July 3}
	\\
The year 1597 saw another unfavorable return to perihelion for comet 12P, similar to or even worse than the one of 1740.  There is also no candidate record in historic sources.
\pagebreak
	\\
		
	\textbf{Perihelion 1527 March 12}
	\\
In November 1526, the comet might have become brighter than magnitude 10 at an elongation of just below $70^{\circ}$.  It continued to perihelion in March	1527, which was reached at an elongation of just above $30^{\circ}$ and with a brightness of perhaps magnitude 4.  There are historic records of comets in 1523 and 1529, but the descriptions do not fit.
	\\
		
	\textbf{Perihelion 1313 May 1}
	\\
The next perihelion before 1385 occurred in 1313, and it was again an unfavorable apparition for earth-based viewers (comparable to 1668). The comet remained at low elongations, and perihelion was attained with magnitude perhaps 4 at elongations below $15^{\circ}$.
	
There was a comet seen on 1313 April 13, about 1.5 months prior to perihelion passage for 12P (when it would have been perhaps at magnitude 5 and at $15^{\circ}$ elongation).  Park and Chae have suggested this comet as a candidate for 12P, too.  Unfortunately, the indicated position in Gemini is not consistent with the position in Aries given by our orbit. So this object can be clearly ruled out.
	
Pingre (p.\ 425) gives a description of the same or another object following the Asian account based on the manuscripts of the historian Mussati (1727, p.\ 554) who lived from 1261-1329: 
	
``In Europe, on April 16, Jupiter and Venus were in conjunction in the sign of Gemini. Four days later a comet was seen in Italy towards the place in the sky where the Sun appeared, when it was about to enter the waters of the ocean: its hairy tail, similar to a whitish smoke, extended to the distance of twenty feet on the west side (it should be read, on the east side). After gradually weakening for a fortnight, this Comet finally vanished. Other Historians similarly testify that the Comet was seen from the west side; therefore his tail could not look to the West."
	
On that date - April 17 - 12P/Pons-Brooks would have been at an elongation of only $13^{\circ}$	and visible low above the western horizon in twilight with a magnitude of maybe 4.5. If the comet would have experienced an outburst around that time it might have been visible even under such conditions but the descritption of a long tail seems to contradict a recent outburst. Probably this account also relates to the Asian Gemini object.
	\\
		
	\textbf{Perihelion 1241 April 11}
	\\
Another unfavourable apparition with the comet becoming brighter than magnitude 6 already at a small elongation of below $35^{\circ}$. Maximum brightness of about magnitude 4.5 was attained at an elongation of $20^{\circ}$. In May the comet had 	traveled southward and became fainter than magnitude 6 at an elongation of 	about $45^{\circ}$. The Japanese text {\it Dai Nihon shi} reports that on February 17, ``a broom star was seen".(Pankenier 2008, p.\ 149). At that time the comet might have been as bright as magnitude 6.5-7 and at an elongation of $43^{\circ}$. It should have been visible only if there was an outburst.
	\\
		
	\textbf{Perihelion 959 January 24}
	\\
The 959 return to perihelion of 12P is similar to that of 1457, when Toscanelli saw the comet from Italy.  There is one comet in historical records in 959, but the details are very uncertain.  They come from a Byzantine text dated 990 and provide no observational details, but rather relate it to the death of Constantine VII Porphyrogenitus (who died on 959 November 9; Kronk 1999).  The comet would then be expected to be bright in January.  Hasegawa (1980) gives a date of 959 Oct. 17 for this	comet and lists another for 959 May, seen from Arabia.
	
Stryuck (1740, p.\ 217) wrote:  ``In the Year 959, a Comet was seen as a dim [literally ``sad and dark"] light.  ({\it Constantin.\ Porphyr.\ incerti Continuat.}, p.\ 289 [e.g., cf.\ Niebuhr 1838]; {\it Symeon Magist \& Logoth. Annal.}, p.\ 496).  When the Comet was seen at the death of the mentioned Emperor, than it	must have appeared in the middle of November."  Struyck also suggested	identify with the 1652 comet:  ``This was the Comet	that was seen in the Year 1652."  Chambers (1889, p.\ 572) cites two sources for a comet in 959, one saying ``a gloomy and obscure star"	(citing the extension of the {\it Chronographia} of Theophanes the	Confessor by Constantine VII, likely taking his citation directly from Struyck) and the other saying that it appeared from Oct.\ 17 to Nov.\ 1 [but a careful reading of the second source, Tackio (1653), doesn't appear to mention either the 959 comet of these specific dates].
	\\
		
	\textbf{Perihelion 886 April 2}
	\\
This apparition is comparable to that of 1241. The comet was already at an elongation of below $40^{\circ}$ when it became brighter than magnitude 6 at the	end of February. Maximum magnitude of 4.5 was attained at an elongation of $22^{\circ}$. It then moves southward and became fainter than magnitude 6 in mid-May at an elongation of about $45^{\circ}$. Three Chinese texts mention a comet seen between June 6 and July 5 Pankenier 2008, p.\ 102). The  {\it Xin Tang shu: Xizong ji}, the {\it Xin Tang shu: tianwen zhi} and the {\it Jiu Tang shu: Xizong ji} say that a ``star became fuzzy" in JI (lunar mansion 7, near $\gamma$ Sgr) and WEI (lunar mansion 6, near $\alpha$ Peg). It then passed through BEIDOU (near $\alpha$ UMa) and SHETI (near $\o$ and $\eta$ Boo). This can not be 12P since it was already situated far south.
	\\
		
	\textbf{Perihelion 813 August 6}
	\\
The 813 return to perihelion was not perfect concerning the geometrical	conditions.  The comet may have become brighter than magnitude 6 in July	at an elongation of about $30^{\circ}$.  Maximum brightness with magnitude perhaps 4 was attained at the beginning of August with a similar elongation.  The comet then moved southward and became fainter than magnitude 6 in September.
	
Interestingly, Pingre (1783, pp.\ 337-338) lists a comet for 813 August 4, but his description (based on the medieval author Theophanes the Confessor) leaves great doubt whether this was indeed a comet:  ``On August 4 a comet was seen, which resembled two moons joined together; they separated, and having taken different forms, at length appeared like a man without a head" (translation from Chambers 1889, p.\ 568). The description sounds more like that of a short-lived transient such as a bright meteor/fireball.  This object of Theophanes was not included in the catalogues of Williams (1871), Ho (1962), and Hasegawa (1980), and is probably
not related to 12P.
	\\
		
	\textbf{Perihelion 742 October 28}
	\\
The 742 return to perihelion was quite favorable. At the beginning of September, comet 12P would be expected to have become brighter than magnitude 6, while	being at an elongation of $65{\circ}$ and a declination of about $+62^{\circ}$. It then moved southward and attained a maximum magnitude of perhaps 1.5 in October, then with an elongation of $50^{\circ}$-$55^{\circ}$ and moving in declination	from $0^{\circ}$ to $-20^{\circ}$ around its closest approach to the earth.
	
Despite these favorable observing conditions, no historic object can be	identified unambiguously from historical records.  Pingre (1783, p.\ 336)	cites several sources for objects around this year.	For 742 and 743, Cedreni (1647, pp.\ 460-461) mentions ``a sign in the sky" and ``a sign in the sky appearing towards the North, which fell down to the ground like dust", respectively. For 743, Hoyland (2011, p.\ 242) gives four different chronicles describing a ``sign in the sky" which more or less agree with each other probably due to copying. This sign is said to have appeared in June and looked like three ``columns of fire that flickered and then remained constant". This sounds very much like an aurora. In June comet 12P would have been at perhaps magnitude 10. An identity, if real at all, is very unlikely.
	
Two of the chronicles go on to say that another such sign was seen in September. Here the month would be matching with the visibility of 12P but details are too scarce to suspect it is a misdated description of a comet.
	
Pingre, Cedreni and Hoyland mention another comet seen from Syria in 744 or 745, possibly in January. This object may have also been seen from Asia (Ho, p.\ 171) Lubienietz says that in 745 a comet was seen in Cancer according to an anonymous report from the German city of Nuremberg. The comet was seen for 39 days. All these objects - if real - have probably no relation to comet 12P.
	\\
		
	\textbf{Perihelion 673 January 8}
	\\
The 673 return to perihelion of comet 12P geometrically falls between the favorable apparitions of 1385 and 1497.  The comet would be predicted to have become brighter than magnitude 6 at the beginning of November 672,	at an elongation of around $75^{\circ}$.  Being at almost $+50^{\circ}$ declination, it then moved southerly, reaching maximum brightness of perhaps magnitude 2 at the end of the year. The elongation was then around $50^{\circ}$-$55^{\circ}$, and the declination around $0^{\circ}$. The comet then continued to move southward and should have become fainter than magnitude 6 in March 673.

Pingre (1783, p.\ 331) lists a comet for this year -- however, with no details that help to decide on any identity with comet 12P/Pons-Brooks.  At first,	he cites two sentences from ancient chronicles:  ``In the first year of	Thierry's reign, we saw a Comet. A fire appeared in the sky for ten days. An extraordinary iris caused so much fright, that it was believed that the last day was near."  The king mentioned was Theoderic III, who became king of Neustria in 673 and king of Austrasia (and thus of all 	Franks) in 679.  Pingre then concludes:  ``All this may be reduced to an aurora borealis.  Of the Authors quoted [...], only one calls it a comet; while he is contemporary, the word comet is sometimes very ambiguous."
	
Stryuck (1740, p.\ 209) wrote:  ``In the Year 673, in the Month of March, a Fire shined 10 days in the Sky. ({\it Centuria. Magdeburg.}, cent.\ 7, cap.\ 13, p.\ 564)"; he also suggested identity of the comet	of 673 with the 1337 and 1558 comets Stryuck (1753, pp.\ 19-20). Hevelius (1668, p.\ 812) and Funccius (p.\ 124) also mention a comet seen for 10 days in 673.
	
Lubinietz (1667, p.\ 116) gives a comet for 674 and refers to Alstedius  (1650, p.\ 506) and Berckringeri (1665, p.\ 32). The two latter sources are based on	de Cesarea (1483) who also gives the phrase with the fire in the skies for ten days.
	
Asian sources do not help in this case.  For the period 672 September 27 to October 25, the Korean \textit{Samguk sagi} and \textit{Jeungbo munheon bigo} speak of a ``broom star" that ``emerged seven times in the north" (Pankenier et al. 2008, p.\ 74; Ho 1962).  This could have been comet 12P if it had been unusually	bright and experiencing an outburst.
	\\
		
	\textbf{Perihelion 386 September 8}
	\\
The return to perihelion in 386 would have produced a quite favorable apparition.  With the end of July, the comet would have appeared in	the morning sky at about magnitude 6 at an elongation of $46^{\circ}$.At perihelion, the comet would have reached a maximum magnitude of about 3.5 at an elongation of $44^{\circ}$; it then started to move southward and would be presumably fainter than magnitude 6 by the end of October at an elongation of $45^{\circ}$.
	
Pingre (1783, p.\ 303) reports a comet seen in this year in Sagittarius but	says that it was a misdated account of comet C/390 Q1.  Ho (1962) reports Asian records that state that there was a comet seen from April/May and disappeared in July/August and situated in Sagittarius.  Biot (1843b) noted a comet seen in China in Sgr in April that was visible until July (see also Carl 1864, p.\ 18).  Hasegawa (1980)	considered this to be a nova.  Since the time and position do not match, an identity with 12P is impossible.
	\\
		
	\textbf{Perihelion 245 September 15}
	\\
A comet observed in the year 245 is probably an earlier sighting of comet 12P/Pons-Brooks.  Pankenier et al. (2008, p.\ 40) provides:  ``6th year of the Zhengshui reign period of King Qi of Wei, 8th month, day wuwu [55]; a white broom star 2 chi long appeared in QIXING [LM 25]. It advanced as far as ZHANG [LM 26] for 23 days in total then was extinguished.	[\textit{Song shu: tianwen zhi}] ch. 23".
	
Other authors (Kronk 1999; Ho 1962; Williams 1871; Pingre 1783) use a slightly different wording as shown as an example in the report by Ho: ``On a wu-wu day in the eighth month of the sixth year of the Cheng-Shih reign-period a white (hui) comet measuring 2 ft (chhih) appeared at the Chhi-Hsing (25th lunar mansion) moving towards the Chang (26th lunar	mansion) and disappeared after 23 days."
	
This means that a comet appeared on 245 Sept. 18 (probably 17.9 UT)	close to $\alpha$, $\iota$, and $\tau$ Hya (QIXING or Chhi-Hsing), and moved towards $\kappa$, $\lambda$, $\mu$, and $\nu$ Hya (ZHANG or Chang).  The tail was about $3^{\circ}$ long.
	
The following derived position assumes that the comet was situated within QIXING.
	\\
	
	245 Sep 17.9: 09h 35m $-05^{\circ}$ 00' (2000.0).
	\\
	
Using Kobayashi's orbit for that epoch, the comet is situated about $8^{\circ}$ from the above position on 245 Sept. 17.9.  Adjusting the perihelion time by +2.9 day QIXING would be around $6.5^{\circ}$ (cf. fig. 5). The brightness would have been at	magnitude 2-3 around that time.\\

\includegraphics[width=0.9\textwidth]{"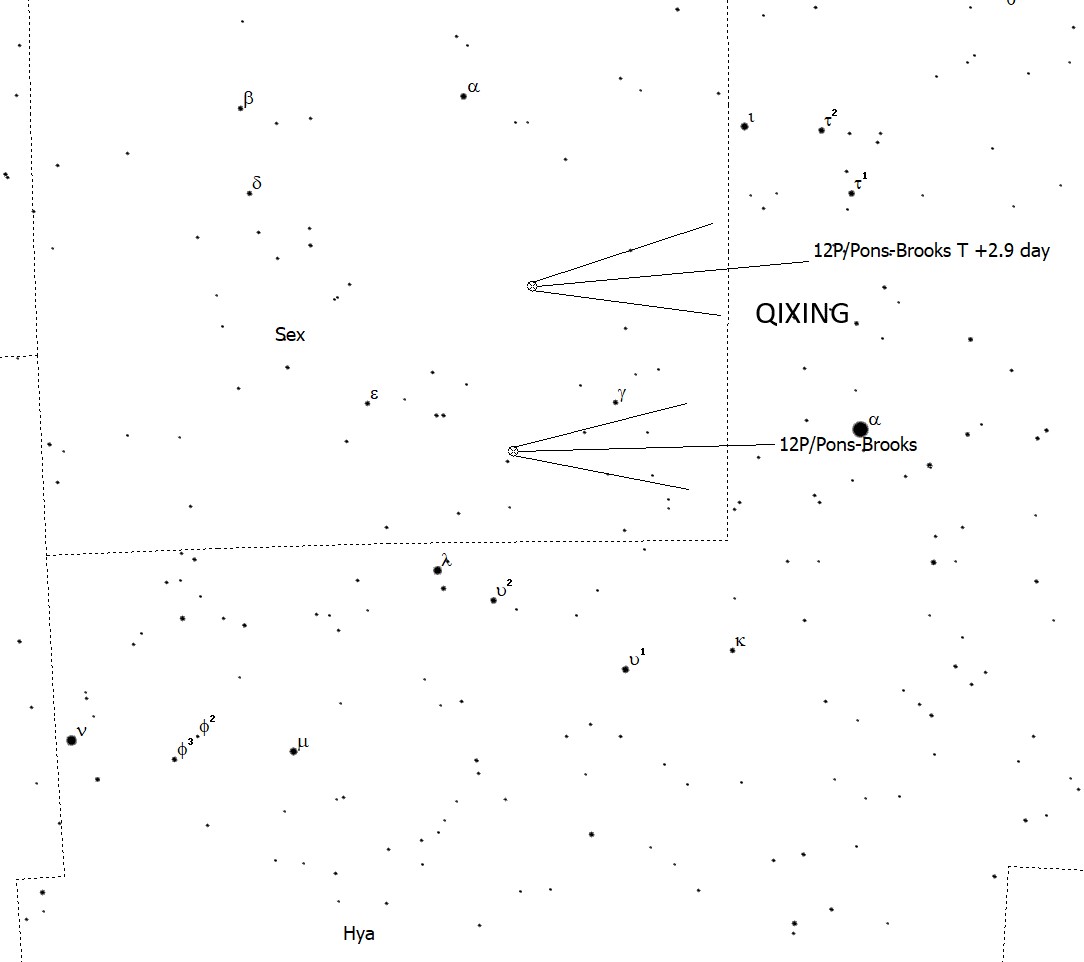"}\\

Figure 5: Position of comet 12P/Pons-Brooks in relation to the Chinese constellation QIXING where the comet was seen on 245 Sep. 17.9 UT.  Shown are the position based on Kobayashi's nominal linked orbit and after an adjustment of t + 2.9 days\\
	
The general direction of movement then carries the comet indeed in the direction of $\kappa$, $\lambda$, $\mu$, and $\nu$ Hya (also around $7^{\circ}$ distance).  The problem here is that Pankenier et al. say that the comet disappeared after 23 days in the region of ZHANG.  This should not have been the case.  After 23 days (Oct. 10), it would already be in Cen, some $23^{\circ}$ away from ZHANG (or even farther, when using the perihelion date of Sept. 9).  All other sources do not connect ZHANG with the date of the last sighting and can be understood	as a scenario in which the comet was moving in the direction of ZHANG and disappeared after 23 days, which would agree with the expected path of comet	12P/Pons-Brooks.
	
Apparently the original text contains some ambiguity in interpretation, and it can indeed be put in both ways.  Upon the request from the authors of this paper,	a word-by-word translation by Ye (2020b) gives:  ``...advanced and arrived Zhang and settled for 23 days and then extinguished."  From a linguistical point-of-view,	it is not fully clear whether the word `settled' refers to the apparition or to the position of the comet with respect to ZHANG.  We nevertheless think that the identification of 12P/Pons-Brooks with the comet of 245 is highly probable and would make it the comet with the third-longest observational arc after 1P/Halley and 109P/Swift-Tuttle.

\section{The apparition of 2024}
	
The impending apparition of 2024 is not very favorable, but will be	better than the last one in 1954.  An analysis by renowned visual observer	Max Beyer (1958) included 76 observations made by himself using the 26-cm equatorial of the Hamburg-Bergedorf observatory (Germany) and shows	numerous outbursts; he notes that the amplitude of these outbursts decreases with decreasing distance from the sun.  At least five outbursts with amplitudes of at least $1^{mag}$ magnitude can be seen in his combined lightcurve
of his visual observations and of photographic ones by G. van Biesbroeck. He finally gives lightcurve parameters of H0 =$4.66^{mag}$ and n = 4.33.
	
This generally agrees with parameters derived by Green (2020) from observations	in the database of the ICQ (H0 = $4.0^{mag}$, n = 3.2), which also roughly agree with the limited brightness information for the apparitions of the 19th century.  An analysis of historic brightness information in Kronk (2003)	confirmed not only the tendency for outbursts but also showed that the comet shows a rather steep decline in brightness after perihelion, hinting to a lightcurve asymmetry.  It should also be taken into account that the	comet has never been observed at distances farther way than 4.5 AU from the	sun pre-perihelion and 2.2 AU post-perihelion. Especially for the pre-perihelion behavior, predictions are very complicated.
	
A new analysis of the apparition of 1954 using data from the ICQ database and including the observations from Beyer show a clear difference between the pre- and the post-perihelion parts of the lightcurve (fig. 6). \\
\begin{verbatim}	
	pre-perihelion: 	H0 = 4.5, n = 5.2
	post-perihelion:	H0 = 5.2, n = 5.1
	
\end{verbatim}
It should be noted, however, that all post-perihelion estimates come from one observer only (A. Jones) and it is possible that the difference between pre- and	post-perihelion is due to observer bias. The lightcurve also shows that the tendency for small outbursts is much more apparent prior perihelion.\\

	\includegraphics[width=1.0\textwidth]{"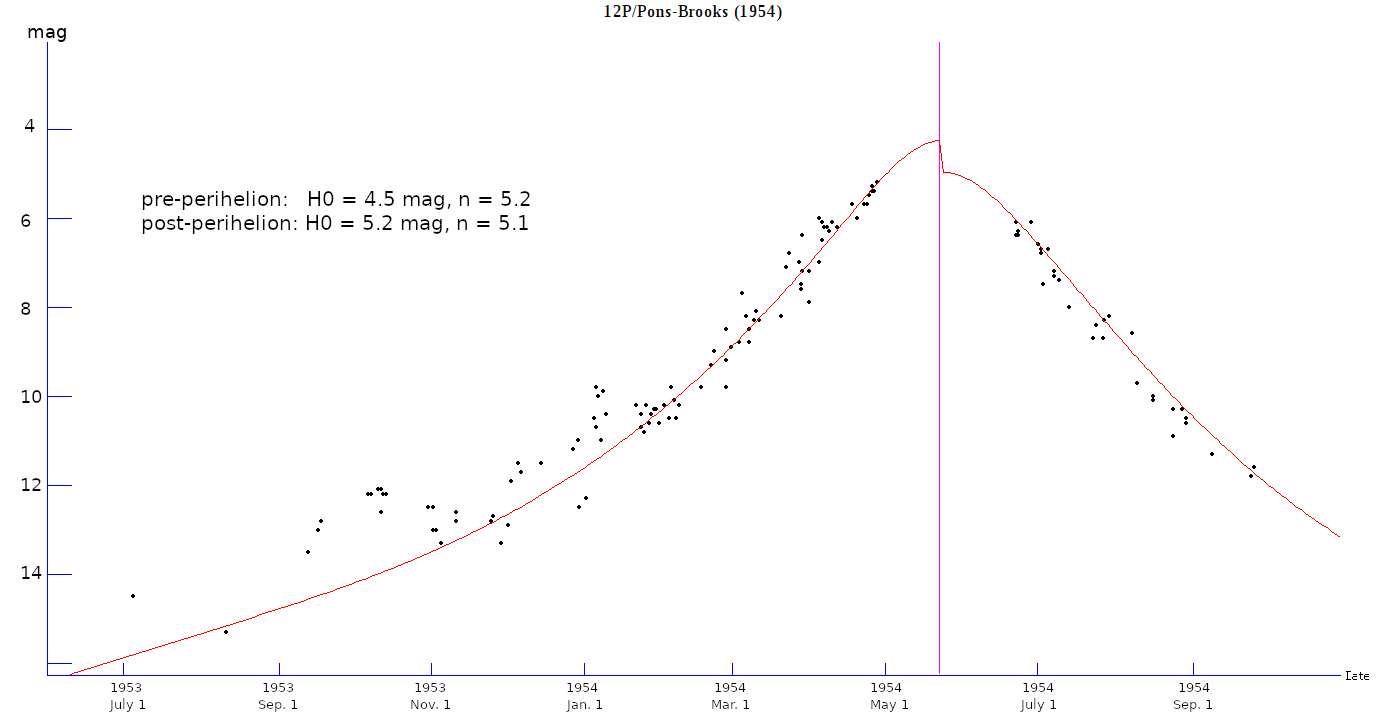"}\\

Figure 6: Lightcurve of comet 12P/Pons-Brooks based on the ICQ database and Max Beyer's	visual observations (1958). (Lightcurve prepared with \textit{``Comet for Windows"}, URL {\tt http://http://www.aerith.net/project/comet.html})\\
	
The ephemeris below uses the values of Green (labelled Mag(a)) and the pre-perihelion parameters from the new analysis (labelled Mag(b)).
\\
\begin{small}
\begin{verbatim}
	Date    TT    R. A. (2000) Decl.     Delta      r     Elong.  Phase  Mag(a) Mag(b)
	2020 03 12    18 17.81   +16 22.8   12.557   12.411    79.3     4.5  18.2   24.2
	2020 04 21    18 17.16   +18 44.6   11.841   12.200   108.7     4.5  18.1   24.0
	2020 05 31    18 07.69   +20 37.1   11.290   11.987   131.5     3.6  17.9   23.8
	2020 07 10    17 53.70   +21 05.0   11.064   11.770   132.1     3.7  17.8   23.6
	2020 08 19    17 42.81   +19 59.5   11.163   11.550   110.1     4.7  17.7   23.6
	2020 09 28    17 40.62   +18 06.7   11.430   11.328    81.6     5.0  17.7   23.5
	2020 11 07    17 47.72   +16 29.6   11.643   11.102    54.9     4.2  17.7   23.4
	2020 12 17    18 01.02   +15 55.5   11.613   10.873    39.6     3.3  17.6   23.3
	2021 01 26    18 15.65   +16 47.2   11.259   10.640    49.1     4.0  17.5   23.1
	2021 03 07    18 26.20   +19 01.2   10.636   10.404    73.9     5.3  17.3   22.9
	2021 04 16    18 27.77   +22 03.7    9.913   10.164   101.7     5.5  17.0   22.6
	2021 05 26    18 18.08   +24 46.4    9.308    9.920   124.7     4.8  16.8   22.3
	2021 07 05    18 00.88   +25 49.0    8.993    9.672   129.6     4.6  16.7   22.1
	2021 08 14    17 45.81   +24 42.2    8.994    9.420   112.0     5.7  16.6   21.9
	2021 09 23    17 41.22   +22 19.7    9.183    9.163    85.7     6.3  16.5   21.8
	2021 11 02    17 48.69   +20 07.2    9.352    8.902    60.3     5.6  16.5   21.7
	2021 12 12    18 04.79   +19 10.3    9.318    8.635    43.9     4.5  16.3   21.5
	2022 01 21    18 23.98   +20 03.5    8.991    8.364    47.9     5.0  16.1   21.3
	2022 03 02    18 39.94   +22 52.2    8.402    8.086    68.2     6.5  15.9   20.9
	2022 04 11    18 46.04   +27 07.4    7.685    7.803    93.1     7.4  15.6   20.5
	2022 05 21    18 36.94   +31 29.9    7.035    7.514   114.7     7.0  15.2   20.1
	2022 06 30    18 13.92   +33 53.8    6.618    7.218   122.7     6.8  15.0   19.8
	2022 08 09    17 50.09   +32 56.6    6.487    6.915   110.9     7.9  14.8   19.5
	2022 09 18    17 40.46   +29 38.9    6.546    6.604    88.9     8.8  14.6   19.2
	2022 10 28    17 48.98   +26 16.1    6.614    6.284    66.5     8.3  14.5   19.0
	2022 12 07    18 11.31   +24 33.7    6.530    5.956    50.8     7.4  14.3   18.6
	2023 01 16    18 40.85   +25 30.4    6.208    5.618    49.4     7.6  14.0   18.2
	2023 02 25    19 10.55   +29 32.8    5.663    5.268    61.8     9.5  13.5   17.6
	2023 04 06    19 32.17   +36 38.0    4.991    4.907    79.4    11.6  13.0   17.0
	2023 05 16    19 33.79   +45 46.6    4.332    4.532    94.9    12.8  12.4   16.2
	2023 06 25    18 59.15   +53 48.6    3.807    4.141   102.1    13.9  11.8   15.4
	2023 08 04    17 55.68   +55 27.5    3.456    3.734    97.9    15.6  11.3   14.6
	2023 09 13    17 19.54   +50 07.6    3.211    3.306    86.5    17.7  10.7   13.8
	2023 10 23    17 33.16   +43 09.6    2.952    2.854    74.6    19.6  10.0   12.8
	2023 12 02    18 27.92   +38 36.5    2.591    2.376    66.4    22.4   9.1   11.5
	2024 01 11    20 06.47   +37 53.3    2.143    1.868    60.6    27.3   7.8    9.7
	2024 02 20    22 48.67   +37 21.1    1.751    1.339    49.5    34.1   6.2    7.4
	2024 03 31    02 05.11   +23 33.2    1.611    0.875    28.6    33.1   4.6    4.8
	2024 05 10    04 31.52   -03 41.3    1.577    0.859    29.6    35.5   4.5    4.6
	2024 06 19    07 00.37   -29 55.0    1.578    1.313    55.9    39.9   5.9    7.0
	2024 07 29    10 07.58   -44 45.7    1.983    1.842    67.0    30.5   7.6    9.4
	2024 09 07    12 40.01   -47 29.2    2.730    2.351    57.7    21.2   9.2   11.5
	2024 10 17    14 23.46   -47 30.8    3.520    2.831    40.0    13.1  10.3   13.1
	2024 11 26    15 39.66   -47 42.0    4.128    3.283    27.4     7.9  11.2   14.3
	2025 01 05    16 36.61   -48 20.5    4.428    3.712    38.8     9.5  11.8   15.1
\end{verbatim}		
\end{small}	
\section{References}
	
	Alstedius, J. H. (1650).  {\it Thesaurus Chronologiae} (Herbornae).\\
	Berckringeri, D. (1665).  {\it Dissertatio Histrorico-Politica de Cometis} (Ultrajectum: Meinard).\\
	Beyer, M. (1958).  ``Physische Beobachtungen von Kometen. X",
	{\it A.N.}\  {\bf 284}, 112-128.\\
	Biot, E. C. (1843a).  ``Catalogue Des Com\`etes observ\'ees en
	Chine depuis l'an 1230 jusqu'\`a l'an 1640 de notre \`ere", in
	{\it Connaissance des Temps ... Pour L'An 1846} (Paris:  Bachelier,
	Imprimeur-Libraire du Bureau des Longitudes), Additions, p.\ 57.\\
	Biot, E. C. (1843b).  ``Catalogue Des \'Etoiles extraordinaires
	observ\'ens en Chine depuis les temps anciens jusqu'\`a l'an 1203 de
	notre \`ere", in {\it Connaissance des Temps ... Pour L'An 1846}
	(Paris:  Bachelier, Imprimeur-Libraire du Bureau des Longitudes),
	Additions, p.\ 64.\\
	Carl, P. (1864).  {\it Repertorium der Cometen-Astronomie}
	(Muenchen:  M. Rieger'sche Universitaets-Buchhandlung).\\
	Cedreni, G. (1647).  {\it Compendium Historiarum. Tomus II} (Paris).\\
	Celoria, G. (1884).  ``Comete del 1457", {\it A.N.}\ {\bf 110}, 174.\\
	Celoria, G. (1894).  ``Con un capitolo sui lavori astronomici del Toscanelli",
	in {\it Raccolta di documenti e studi pubblicati dalla R.\ Commissione
		colombiana} (Roma:  Ministero della pubblica istruzione), part.\ 5, vol.\ 2.\\
	Celoria, G. (1921).  ``Sulle osservazioni di comete fatte da Paolo Dal Pozzo
	Toscanelli e sui lavori astronomici in generale", {\it Pubblicazioni del
		Reale Osservatorio Astronomico di Brera in Milano}, 55.\\
	Chambers, G. F. (1889).  {\it A Handbook of Descriptive and Practical
		Astronomy}, 4th ed. (Oxford:  Clarendon Press), Vol.\ 1.\\
	de Cesarea, E. (1483).  {\it Chronicon} (venezia: Erhard Ratdolt).\\
	Festou, M. C.; B. Morando; and P. Rocher (1985).  ``The orbit of periodic
	comet Crommelin between the years 1000 and 2100",
	{\it Astron. Ap.}\ {\bf 142}, 421-429.\\
	Funccio, J. (1578). {\it Chronologia} (Witebergae).\\
	Galle, J. C. (1894).  {\it Verzeichniss der Elemente der bisher
		berechneten Cometenbahnen} (Leipzig:  Verlag von Wilhelm Kugelmann).\\
	Gould, G. P., transl.\ (1977).  {\it Manilius:  Astronomica}
	(Cambridge, MA:  Harvard University Press), p.\ 77.\\
	Green, D. W. E. (2020a). {\it CBET} No. 4727.\\
	Green, D. W. E. (2020b). {\it CBET} No. 4805.\\
	Hasegawa, I. (1979).  ``Orbits of ancient and medieval comets",
	{\it Publ.\ Astron.\ Soc.\ Japan} {\bf 31}, 257-270.\\
	Hasegawa, I. (1980).  ``Catalogue of Ancient and Naked-Eye Comets",
	{\it Vistas Astron.}\ {\bf 24}, 59-102.\\
	Hevelii, J. (1668).  {\it Cometographia, totam naturam cometarum} (Gedani, Simon Reininger).\\
	Hind, J. R. (1846). ``Schreiben des Herrn J. R. Hind an den Herausgeber", 
	{\it A.N.}\ {\bf 23}, 177.\\
	Ho, P. Y. (1962).  ``Ancient and mediaeval observations of comets and
	novae in Chinese sources", {\it Vistas Astron.}\ {\bf 5}, 127-225.\\
	Hoyland, R. G. (2011).  {\it Theophilus of Edessa's Chronicle},
	Translated Texts for Historians, vol. 57 (Liverpool Univ. Press).\\
	Jervis, J. L. (1985).  {\it Cometary Theory in Fifteenth-Century Europe},
	Studia Copernicana (Wroclaw).\\
	Kobayashi, T. (2020). {\it Nakano Note}, No. 4136 (2020 June 28); posted
	at website URL\\
	 {\tt http://www.oaa.gr.jp/\textasciitilde oaacs/nk/nk4048.htm}.\\
	Kronk, G. W. (1999).  {\it Cometography}, Vol.\ 1 (Cambridge University
	Press).\\
	Kronk, G. W. (2003).  {\it Cometography}, Vol.\ 2 (Cambridge University
	Press).\\
	Kronk, G. W. (2007).  {\it Cometography}, Vol.\ 4 (Cambridge University
	Press).\\
	Lubienietz, S. (1667).  {\it Systematis Cometici Tomus Secundus . . . Theatri
		Cometici pars posterior . . . Historia Cometarum, . . .} (Amsterdam:
	Francisco Cuyper).\\
	Marsden, B. G. (1975).  {\it Catalogue of Cometary Orbits}, 2nd ed.
	(Cambridge, MA:  Smithsonian Astrophysical Observatory).\\
	Meyerus Baliolanus, J. M. (1561). {\it Commentarii sive annales rerum 
		Flandricarum} (Anvers: Steelsius). \\
	Mussati, A. (1723). ``Historia Augusta", {\it Rerum Italicarum Scriptores}. 
	Tomus Decimus. (Mediolanum. Flandricarum)\\
    Nakano, S. (2020a). {\it Nakano Note}, No. 4048 (2020 Mar. 3); posted
    at website URL \\
    {\tt http://www.oaa.gr.jp/\textasciitilde oaacs/nk/nk4048.htm}.\\
    Nakano, S. (2020b). {\it Nakano Note}, No. 4136 (2020 Jun 28); posted
    at website URL\\
     {\tt http://www.oaa.gr.jp/\textasciitilde oaacs/nk/nk4136.htm}.\\
    Niebuhr, B. G., ed.\ (1838).  {\it Corpus Scriptorum Historiae Byzantinae.
	Theophanes Continuatus} (Bonn:  Impensis Ed.\ Weberi).\\
    Pankenier, D. W.; Z. Xu; and Y. Jiang (2008).  {\it Archaeoastronomy in
	East Asia} (Amherst, NY:  Cambria Press).\\
    Park, S.-Y.; and J. Chae (2007).  ``Analysis of Korean Historical Comet
    Records", {\it Publ.\ Korean Astron. Soc.}\ {\bf 22}, 151-168.\\
    Peirce, B. (1846). {\it American Almanac for 1847}, p.\ 83.\\
    Pingre, A. G. (1783).  {\it Cometographie ou traite historique et
	theoretique des cometes. Tome premier} (Imprimerie Royale, Paris).\\
    Procter, M.; and A. C. D. Crommelin (1937).  {\it Comets:  Their Nature,
	Origin, and Place in the Science of Astronomy} (The Technical Press, London).\\
    Ramsey, J. T.; and A. L. Licht (1997).  {\it The Comet of 44 B.C.\ and
	Caesar's Funeral Games} (Atlanta, GA:  Scholar's Press), p.\ 94.\\
    Schulhof, L. (1885).  ``Ueber muthmassliche frühere Erscheinungen des
    Cometen 1873 VII", {\it A.N.}\ {\bf 113}, 143-144.\\
    Stephenson, F. R. (1997).  {\it Historical Eclipses and Earth's Rotation}
    (Cambridge University Press), p.\ 515.\\
    Struyck, N. (1740).  ``Inleiding tot de Algemeene Kennis der Comeeten,
    of Staarsterren", in {\it Inleiding tot de Algemeene Geographie, benevens
	eenige Sterrekundige en andere Verhandelingen}  (Amsterdam:  Isaak Tirion).\\
    Struyck, N. (1753).  ``Vervolg van de Beschryving der Comeeten of
    Staartsterren", in {\it Vervolg van de Beschryving der Staartsterren,
	en Nader Ontdekkingen Omtrent den Staat van 't Menschelyk Geslagt}
    (Amsterdam:  Isaak Tirion).\\
    Sueyro, E. (1624). {\it Anales de Flandes, Segunda Parte} (Anvers: Pedro 
    y Iuan Belleros).\\
    Tackio, J. (1653).  {\it Coeli Anomalon, id est, De Cometis, sive Stellis
	Crinitis ...} (Giessen, Germany:  Ex officina Typographica Chemliniana).\\
    Werlauff, E. C. (1847). {\it Íslenzkir annálar, sive Annales Islandici ab 
	anno Christi 803 ad annum 1430}  (sumptibus Legati Arnæ-Magnæani).\\
    Williams, J. (1871).  {\it Observations of Comets from BC 611 to
	AD 1640} (London:  Stangeways and Walden).\\
    Wyttenbach, J. H., Müller, M. F. J. (1838). {\it Gesta Trevirorum integra 
	lectionis varietate et animadversionibus illustrata ac indice duplici 
	instructa. Vol. II} (Trier: Lintz).\\
    Ye, Q. et. al. (2020a).  ``Recovery of Returning Halley-Type Comet 12P/Pons-Brooks with the Lowell Discovery Telescope", {\it RNAAS}\ {\bf 4}, No. 7, 2020 July 7.\\
    Ye, Q. (2020b).  Personal communication.\\

\end{document}